\documentclass[tightenlines,notitlepage,11pt,preprintnumbers]{revtex4-1} 

\usepackage{amsmath}    
\usepackage{graphicx}   
\usepackage{verbatim}   
\usepackage{subfigure}  
\usepackage{hyperref}   
\usepackage{mathtools}
\newcommand{\GeV}{\ensuremath{\mathrm{GeV}}}
\newcommand{\GeVc}{\ensuremath{\mathrm{GeV}}} 
\newcommand{\GeVcc}{\ensuremath{\mathrm{GeV}}}
\newcommand{\TeV}{\ensuremath{\mathrm{TeV}}}
\newcommand{\TeVcc}{\ensuremath{\mathrm{TeV}}}

\begin{document}

\preprint{CMS CR-2014/277}

\title{Beyond standard model Higgs physics: prospects for the High
  Luminosity LHC}
\author{Andr\'e G. Holzner}
\email{Andre.Georg.Holzner@cern.ch}
\affiliation{University of California at San Diego, 9500 Gilman Drive, La
         Jolla, California 92093, USA\\
        \\
         on behalf of the ATLAS and CMS collaborations}

\begin{abstract}
This article summarises a talk given at Higgs Hunting 2014 on projections of 
the sensitivity 
of the ATLAS and CMS experiments 
to beyond
standard model Higgs physics at the High Luminosity Large Hadron
Collider. 
We describe results on vector boson scattering, searches for
additional neutral Higgs bosons in two Higgs doublet models
and flavour changing neutral current top decays.
\end{abstract}

\maketitle

\section{Introduction}

The High Luminosity Large Hadron Collider (HL-LHC) is a major
upgrade of the LHC to deliver an integrated luminosity of
$3000\ \mathrm{fb^{-1}}$ within a decade to both ATLAS and CMS.
One of its main goals is permitting the precise measurement of the
properties of the recently discovered standard model (SM) Higgs like 
boson~\cite{Aad:2012tfa,Chatrchyan:2012ufa}. 

In this report, we highlight a few recent results related
to probing the nature of this boson and sensitivity to 
beyond standard model Higgs physics.
Section~\ref{sec:weak-boson-scattering} discusses the expected
sensitivity to non-standard model Higgs couplings in weak boson
scattering processes. The following section~\ref{sec:2HDM}
describes the current estimates for sensitivity to searches for 
additional neutral Higgs bosons in two Higgs doublet models.
Finally, section~\ref{sec:top-fcnc-decays} reviews the projections
for flavour changing neutral current top decays involving Higgs
bosons and section~\ref{sec:conclusions-outlook} concludes and discusses
the next steps.

\section{Weak boson scattering}\label{sec:weak-boson-scattering}

\begin{figure}[h!]
\centering
\includegraphics[scale=0.5]{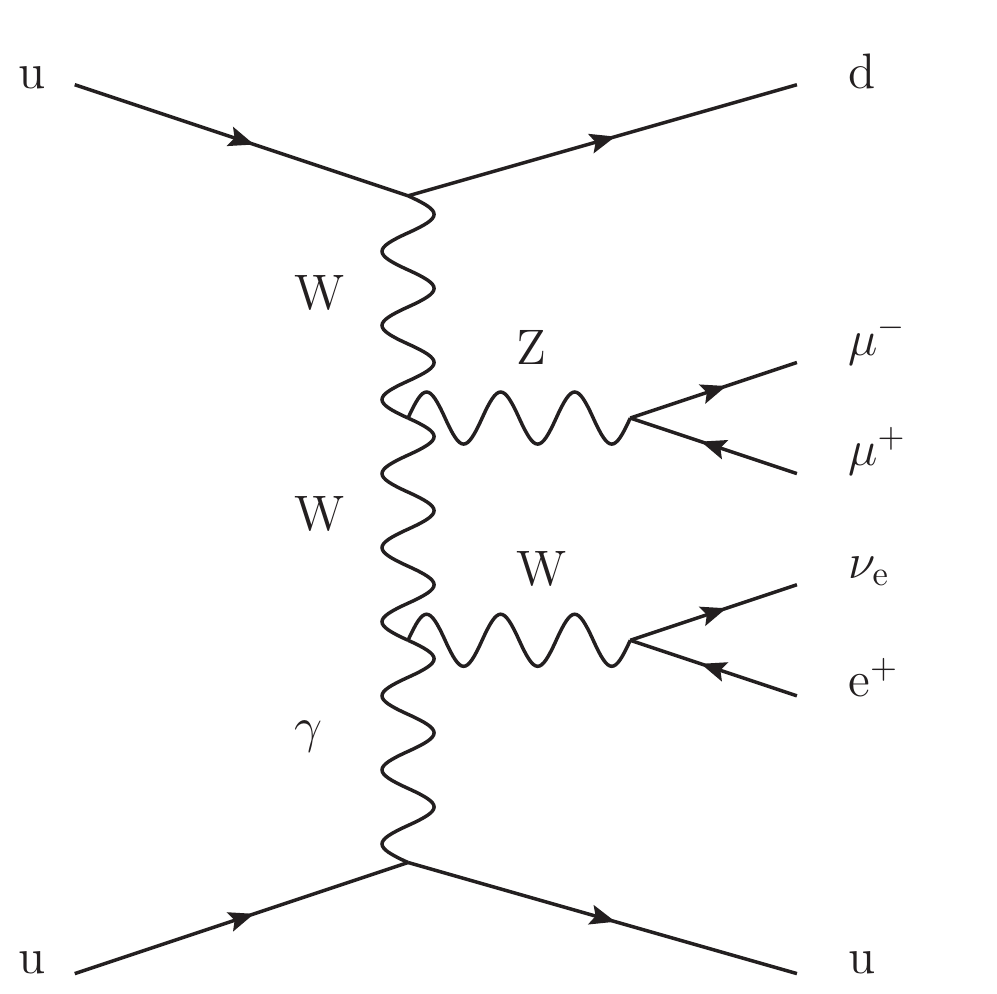}
\includegraphics[scale=0.5]{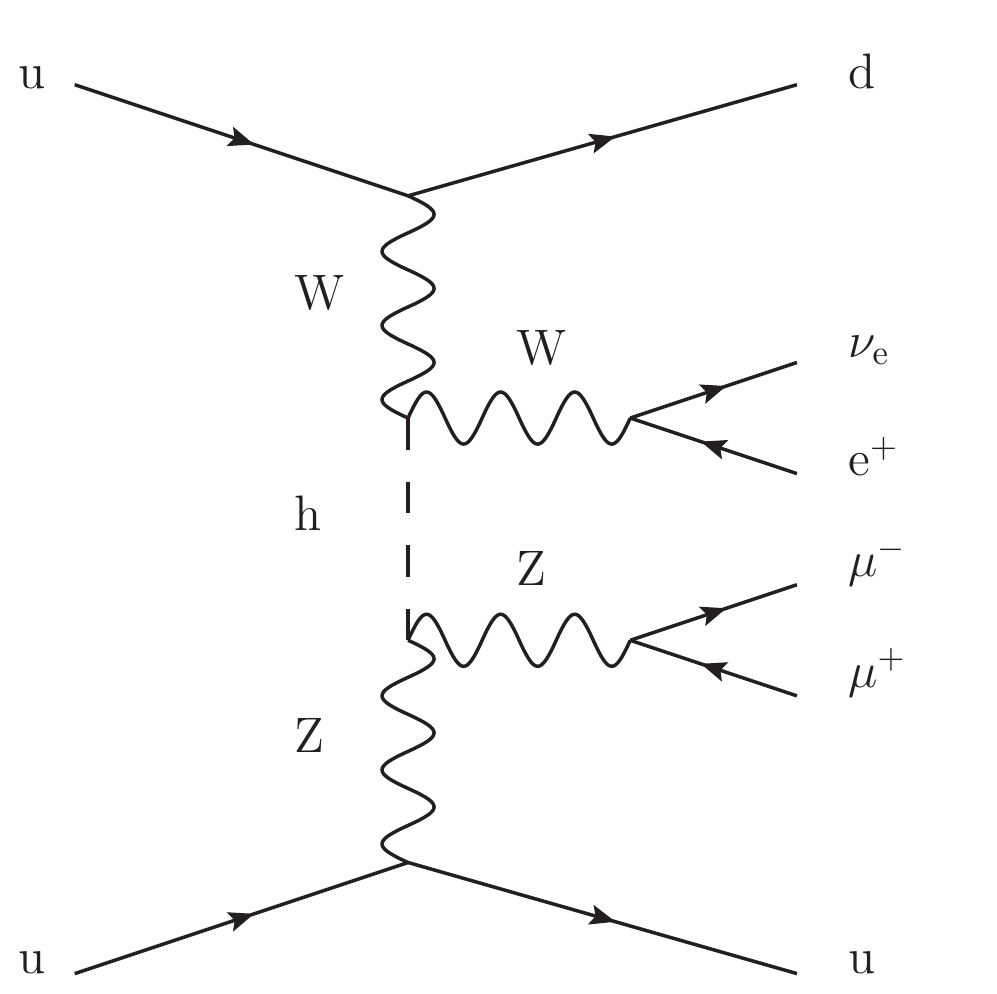}
\includegraphics[scale=0.5]{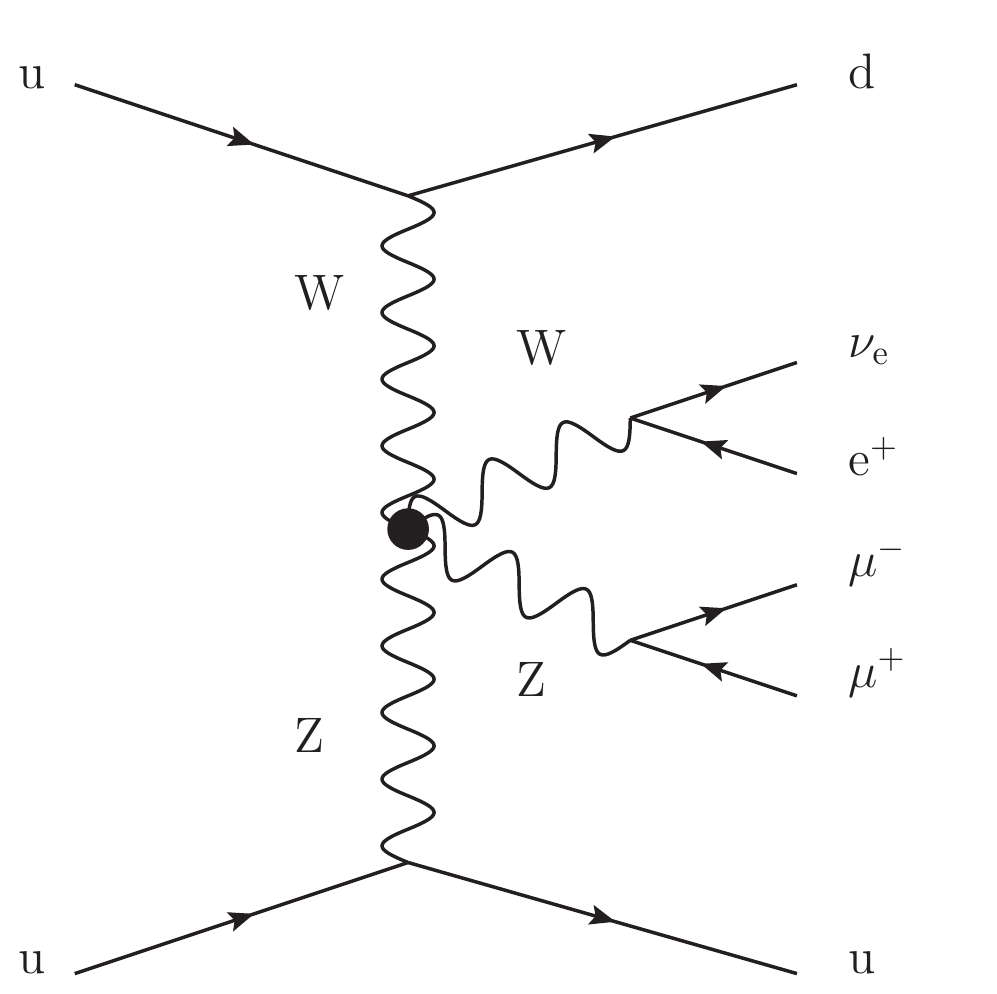}
\caption{\label{fig:vv-scattering-diagrams} Example diagrams related to vector boson scattering at the LHC
for a given final state.
The left diagram is a vector boson scattering process, 
the middle diagram shows a process involving the standard model Higgs boson 
and the right diagram corresponds to anomalous quartic gauge couplings.
 }
\end{figure}

In addition to providing a mechanism for electroweak symmetry breaking, 
the standard model Higgs boson also restores the unitarity of scattering
of weak bosons such as $\mathrm{W^+} \mathrm{W^-} \rightarrow \mathrm{W^+}
\mathrm{W^-}$. 
Figure~\ref{fig:vv-scattering-diagrams} shows examples
of Feynman diagrams involving vector boson scattering in proton proton
collisions. 
The first diagram has only vector bosons 
in the core interaction while the second diagram (interfering negatively with diagrams
of the first type) involves a Higgs boson. 
The third diagram
illustrates how anomalous quartic vector boson couplings can contribute
to the same final state.
Due to the strong interference of the first and second diagrams, any 
beyond the standard model processes like the third diagram are expected
to significantly modify the kinematic distributions of the final state particles.

Comparing the middle and right diagrams it becomes also apparent that 
contributions from new heavy bosons would first be seen as anomalous
{\sl quartic} rather than triple gauge couplings because such bosons 
contribute to quartic couplings at leading order while the
contribution to triple gauge couplings only affects higher orders~\cite{Yang:2012vv}.

The experimental signature is two jets with a large difference in pseudorapidity,
multiple leptons and in some cases missing transverse energy:


\begin{itemize}

\item {$pp \rightarrow ZZjj \rightarrow \ell^+\ell^-\ell^+\ell^- jj$}

This channel suffers from a small branching ratio, 
but has a very characteristic final state with four charged leptons on the other hand.
The kinematics of the outgoing diboson system is fully reconstructable, 
making it possible to measure the cross section as function of centre of mass energy of
the diboson interaction. The ATLAS selection~\cite{ATL-PHYS-PUB-2013-006} requires four 
charged leptons (electrons or muons) with a transverse momentum above 25~\GeVc\ which must be consistent with two
$\mathrm{Z}$ boson candidates (pairs of opposite charge and same flavour). At least two jets with a transverse momentum above 50 \GeVc{} 
are required and two of them must form a dijet mass above 1 \TeVcc.
%
%
\item {$pp \rightarrow W^\pm Z jj \rightarrow \ell^\pm\nu\ell^+\ell^- jj$}

Cross section times branching ratio is larger in this final state than in the ZZjj channel,
but reconstructing the full kinematics of the diboson system becomes more difficult
due to the presence of a neutrino. 

Both ATLAS~\cite{ATL-PHYS-PUB-2013-006} and CMS~\cite{CMS-PAS-FTR-13-006} have estimated 
up to which energy scale anomalous couplings could be observed at HL-LHC. 
The ATLAS (CMS) event selection requires exactly three high transverse momentum leptons (electrons or muons) 
with $p_T > 25\ (20)\ \GeVc$. Two of them must be consistent a 
$\mathrm{Z}$ boson decay (same flavour and opposite charge, invariant mass close to the $\mathrm{Z}$ boson mass).
At least two jets with a transverse momentum above 50 \GeVc\ must be present with a 
dijet invariant mass of at least 1000 (600) \GeVcc. The CMS analysis additionally requires
that the two jets are separated by at least 4.0 in pseudorapidity.

The ATLAS study makes the assumption that the missing
transverse energy is entirely due to the neutrino and requires that the invariant
mass of the neutrino and the charged lepton not associated to the $\mathrm{Z}$ boson decay 
is equal to the $\mathrm{W}$ boson mass
in order to estimate the longitudinal
component of the neutrino momentum.

Expected distributions of the $\mathrm{WZ}$ (transverse) mass, which are used to discriminate 
anomalous quartic coupling signals from standard model processes, are shown in Figure~\ref{fig:vbs-wz-mass}.


\begin{figure}[h!]
\centering
\includegraphics[width=0.48\textwidth]{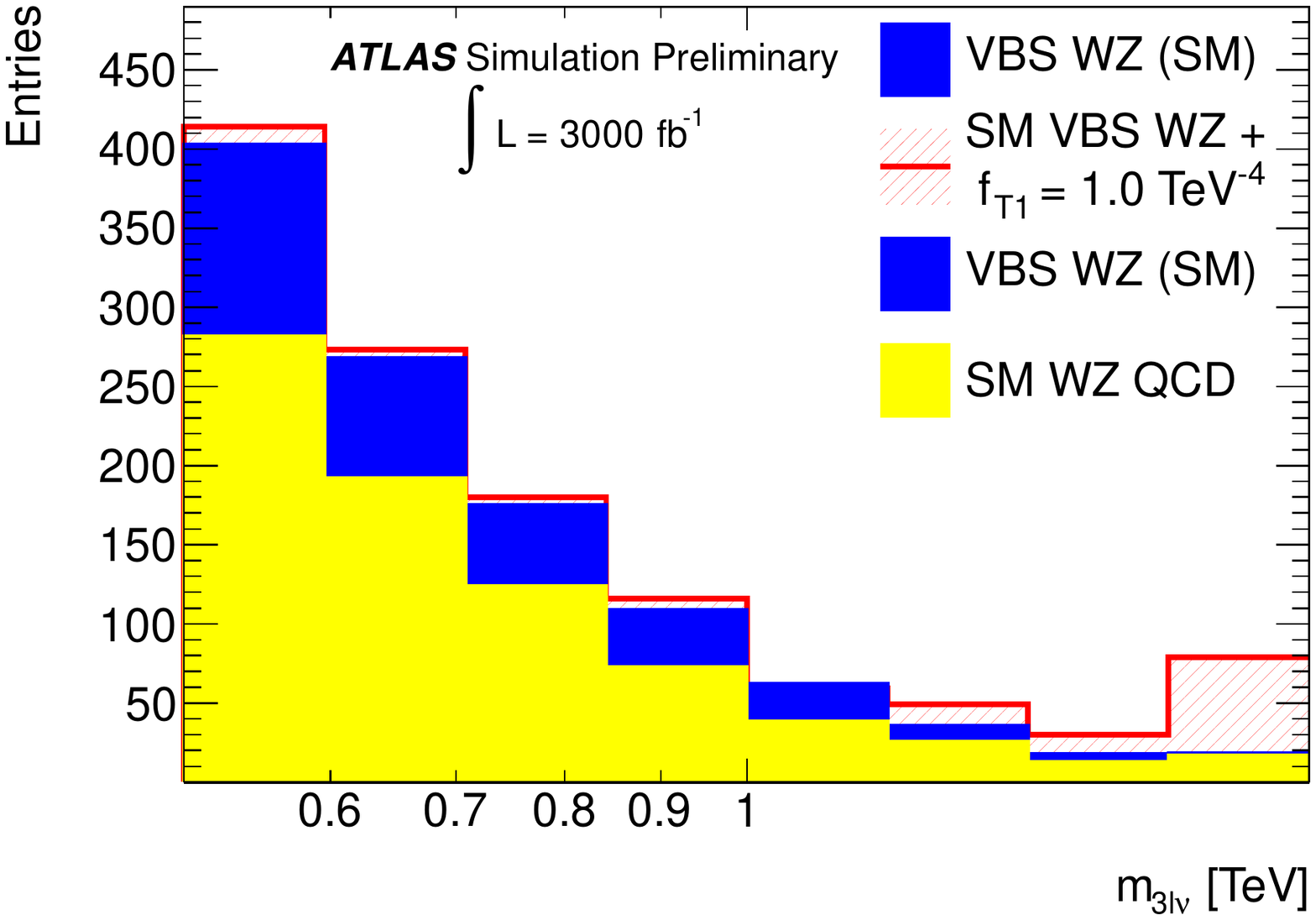}
\includegraphics[width=0.48\textwidth]{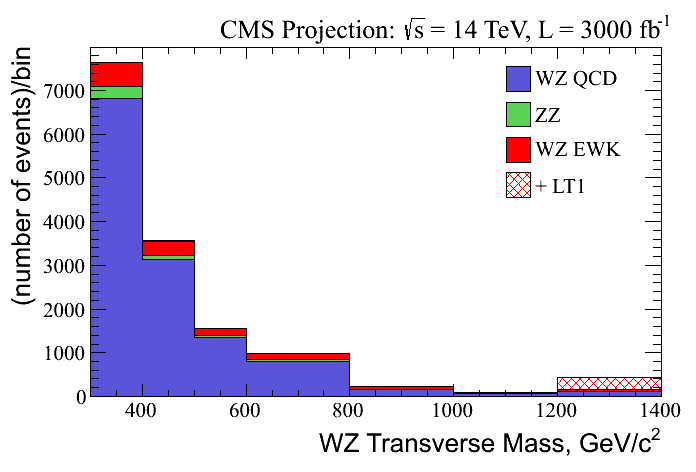}
\caption{\label{fig:vbs-wz-mass} 
$\mathrm{WZ}$ invariant mass distributions in the $pp \rightarrow WZjj \rightarrow \ell^\pm\nu\ell^+\ell^- jj$ channel
for the ATLAS study~\cite{ATL-PHYS-PUB-2013-006} (left) and $\mathrm{WZ}$ transverse mass distribution for CMS~\cite{CMS-PAS-FTR-13-006} (right).}
\end{figure}

\item {$pp \rightarrow W^+W^+jj \rightarrow \ell^+\nu\ell^+\bar{\nu} jj$}

Replacing also the second $\mathrm{Z}$ boson by a $\mathrm{W}$ boson, cross section times branching ratio increases again 
with respect to the previous final states, but the second neutrino 
makes the reconstruction of the full kinematics of the weak boson scattering system impossible. 

The ATLAS experiment has studied the sensitivity at HL-LHC for this channel. Exactly two identified leptons of the {\it same} charge, 
with a transverse momentum above 25~\GeVc\ are required in the selection together with at least 
two identified jets with a transverse momentum above 50~\GeVc. The invariant mass of the
two selected highest $p_T$ jets must exceed 1 \TeVcc. 
The distribution of the mass of the two jets and two leptons is used to discriminate signal from background.

Both ATLAS and CMS experiments have sensitivity 
to this channel already with LHC Run I data~\cite{Aad:2014zda,CMS-PAS-SMP-13-015}.

\end{itemize}

Table~\ref{tab:vbs-sensitivity-summary} summarises the expected sensitivities for these three final states.

\begin{table}[h]
  \centering
\begin{tabular}{|c|c|c|c|c|} \hline
experiment  & final state                   & operator                                & dimension & $5\sigma$ reach for $3000\ \mathrm{fb}^{-1}$ \\ \hline 
ATLAS       & $\ell^+\ell^-\ell^+\ell^-jj$  & ${\cal O}_{\Phi\mathrm{W}}/ \Lambda^2$  & 6         & $16\ \TeV^{-2}$ \\ 
ATLAS / CMS & $\ell^\pm\nu\ell^+\ell^-jj$   & ${\cal L}_{T,1} / \Lambda^4$            & 8         & $0.55\ /\ 0.6\ \TeV^{-4}$\\ 
ATLAS       & $\ell^\pm\nu\ell^\pm\nu jj$   & ${\cal L}_{S,0} / \Lambda^4$            & 8         & $4.5\ \TeV^{-4}$ \\
\hline
\end{tabular}
\caption{\label{tab:vbs-sensitivity-summary} Projected $5\sigma$ discovery reach for the vector boson scattering analyses at HL-LHC. For the same strength of the coupling and order of the operator, smaller numbers
correspond to a higher scale $\Lambda$ up to which deviations from the standard model can be seen at $5\sigma$ significance.
The dimension 6 and 8 operators are defined in~\cite{Degrande:2012wf} and \cite{Eboli:2006wa}.
For comparison, limits on the $S_0$ operator coefficient obtained with Run I data are $^{+40}_{-38}$ ($^{+43}_{-42}$) $\TeV^{-4}$~\cite{CMS-PAS-SMP-13-015}.
}
\end{table}
 
\section{Two Higgs Doublet Models}\label{sec:2HDM}

Two Higgs Doublet models (2HDM) are an extension of the standard model where a second
Higgs doublet is added, giving rise to three neutral Higgs bosons $h, H$ and $A$
and two charged Higgs bosons $H^\pm$. The most widely known example of a 2HDM is
the minimal supersymmetric standard model (MSSM).
2HDMs can be parametrised in several ways, a common set of parameters~\cite{Eriksson:2009ws} is the following:
the four masses of the Higgs bosons $m_h$, $m_H$, $m_A$ and $m_{H^\pm}$, 
the ratio of the vacuum expectation values of the two doublets $\tan \beta$, 
$\sin(\beta - \alpha)$ (where $\alpha$ is the mixing angle between the CP-even Higgs bosons)
and
the Higgs potential parameters $m_{12}^2$, $\lambda_6$ and $\lambda_7$. 

Both ATLAS~\cite{ATL-PHYS-PUB-2013-016} and CMS~\cite{CMS-PAS-FTR-13-024} experiments have estimated
their sensitivity to searches for the heavy CP-even Higgs boson H and the CP-odd Higgs boson A
at HL-LHC. 

\subsection{$pp \rightarrow H \rightarrow ZZ \rightarrow \ell^+\ell^-\ell^+\ell^-$ }

The selection in this channel is very similar to the four lepton selection
in the search for the standard model Higgs. ATLAS
has estimated the sensitivity at HL-LHC mainly by
scaling the cross sections of the Run I analysis~\cite{Aad:2013wqa} to $\sqrt{s} = 14\ \TeV$. 
CMS has performed
a study using DELPHES~\cite{deFavereau:2013fsa}, requiring exactly four identified and isolated
leptons (muons or electrons).
The leading (subleading) lepton must have $p_T > 20\ (10)\ \GeVc$ or the leading lepton
must have more than $30\ \GeVc$ transverse momentum, the other two leptons
must have $p_T$ above 10 \GeVc\ (5\ \GeVc\ for muons).
The leptons must be consistent with the production of two Z bosons. 

\begin{figure}[h!]
\centering
\includegraphics[width=0.47\textwidth]{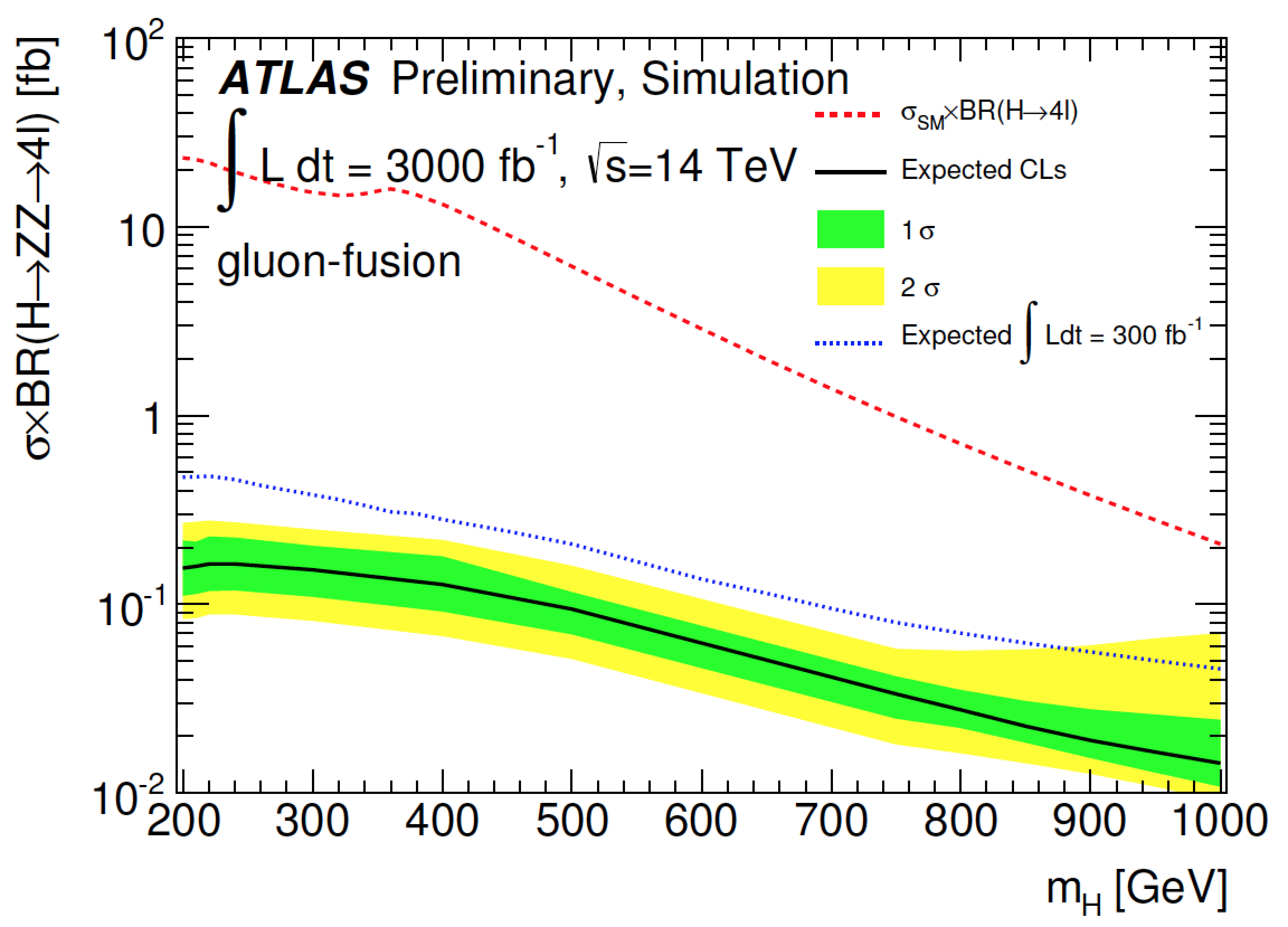}
\raisebox{1.5mm}{\includegraphics[width=0.48\textwidth]{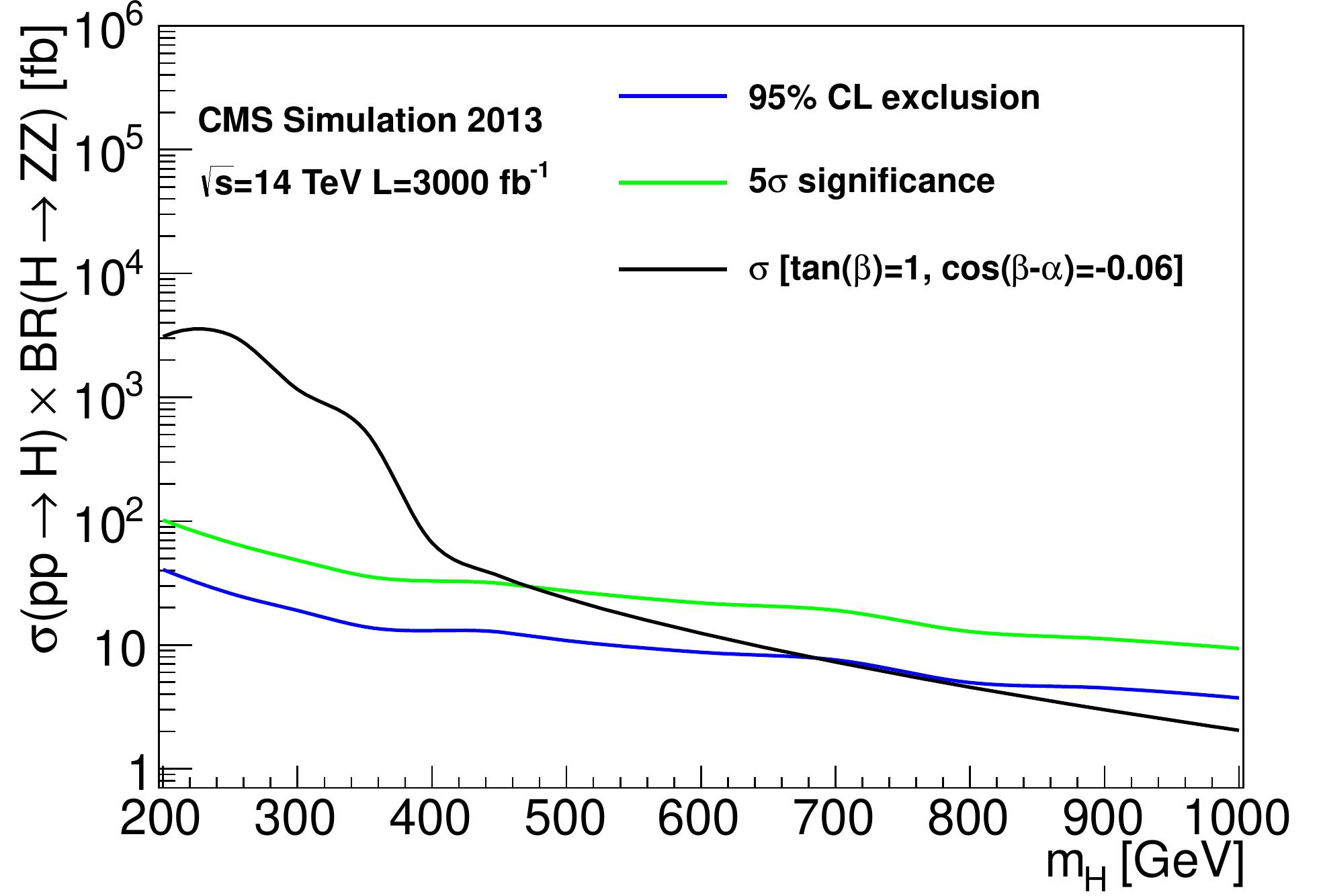}}
\caption{
\label{fig:2hdm-h-zz-llll}
  Exclusion sensitivity as function of the heavy CP-even Higgs mass $m_H$ for the gluon fusion production mechanism,
  for the ATLAS study~\cite{ATL-PHYS-PUB-2013-016} (left) and the CMS study~\cite{CMS-PAS-FTR-13-024} 
  (right; branching ratio $\mathrm{ZZ} \rightarrow 4\ell$ not included).
}
\end{figure}

The expected sensitivity for the gluon fusion process is shown in Figure~\ref{fig:2hdm-h-zz-llll}.


\subsection{$pp \rightarrow A \rightarrow Zh \rightarrow \ell^+\ell^-b\bar{b}$ }

The selections in the search for the pseudoscalar Higgs boson A are very similar for both experiments: two leptons (electrons or muons)
are required (where the CMS study has separate $p_T$ thresholds for the leading
and subleading lepton) which must be consistent with a Z boson decay. 
Two b-jets must be present and must form an invariant mass in a relatively large window around 
the standard model Higgs mass. Both experiments use a b-tag selection corresponding
to a b-jet signal efficiency of 70\%.

In order to further discriminate against backgrounds, the ATLAS study requires
$\Delta R(bb)$ to be within a (slightly $m_A$ dependent) range, while CMS imposes
an upper bound on $\Delta \phi(\ell\ell)$, a lower bound on the transverse momentum
of the dilepton system and requires the ratio of the transverse momenta
$p_T(\mathrm{Z}) / p_T(\mathrm{h})$ to be within a certain range covering one.

To improve the mass resolution for the $A$ boson with respect to the straightforward four body mass,
ATLAS uses the following variable:

\begin{equation*}
m_A^\mathrm{rec} \coloneqq m_{\ell\ell bb} - m_{\ell\ell} - m_{bb} + m_Z + m_h
\end{equation*}

i.e.{} the reconstructed dilepton and dijet masses are replaced by the nominal Z and Higgs masses
$m_Z$ and $m_h$.

\begin{figure}[h!]
\centering

\includegraphics[width=0.48\textwidth]{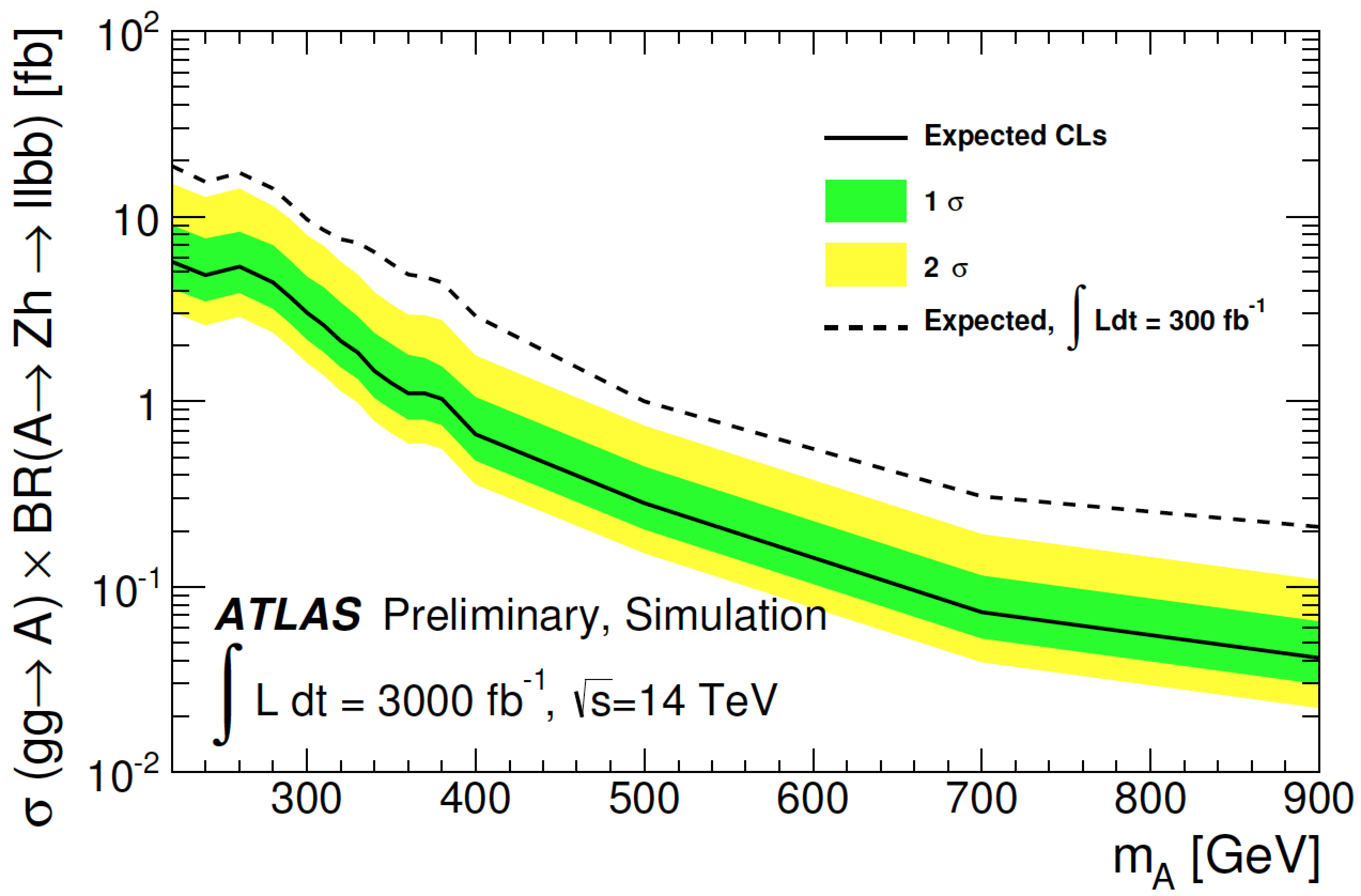}
\raisebox{1mm}{\includegraphics[width=0.455\textwidth]{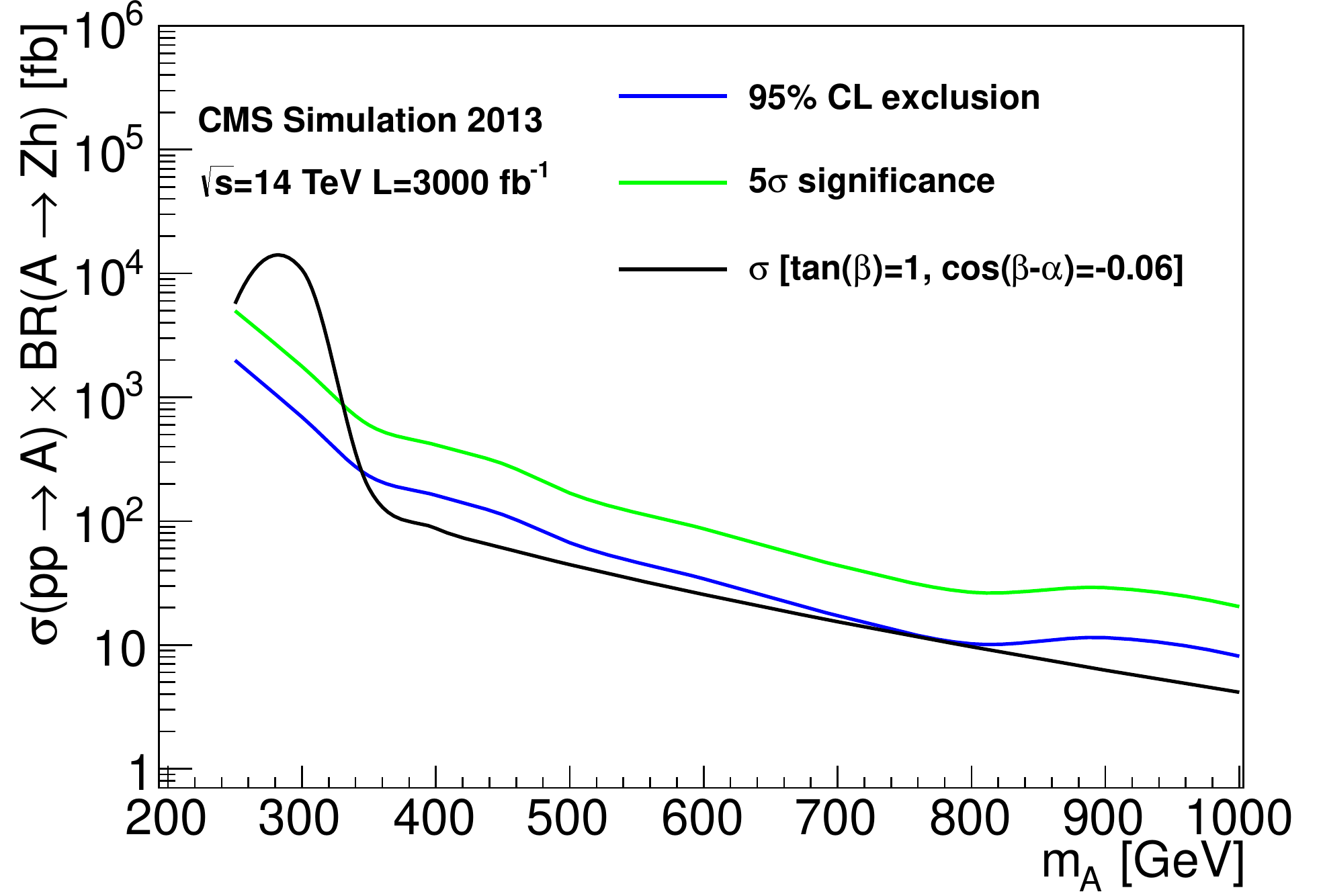}}
\caption{
\label{fig:2hdm-A-hz-bbll}
  Exclusion sensitivity as function of the CP-odd Higgs mass $m_A$,
  for the ATLAS study~\cite{ATL-PHYS-PUB-2013-016} (left) and the CMS study~\cite{CMS-PAS-FTR-13-024} 
  (right; branching ratio $\mathrm{hZ} \rightarrow bb\ell\ell$ not included).
}
\end{figure}

Expected sensitivities for this channel at HL-LHC are shown in Figure~\ref{fig:2hdm-A-hz-bbll}.


\subsection{$pp \rightarrow H/A \rightarrow \mu^+\mu-$ }

At high values of $\tan\beta$, the decays of A and H to $\mu^+\mu^-$ becomes relevant
and is complementary to the $A \rightarrow hZ$ channel. Furthermore, this
channel benefits from an excellent mass resolution.
In the ATLAS study,
exactly two muons with opposite charge are required. 
Events are split into two categories to better exploit the features of the
two dominant production mechanisms gluon fusion (no b-tagged jet allowed)
and b-associated production (at least one b-tagged jet required).

\begin{figure}[h!]
\centering
\includegraphics[width=0.48\textwidth]{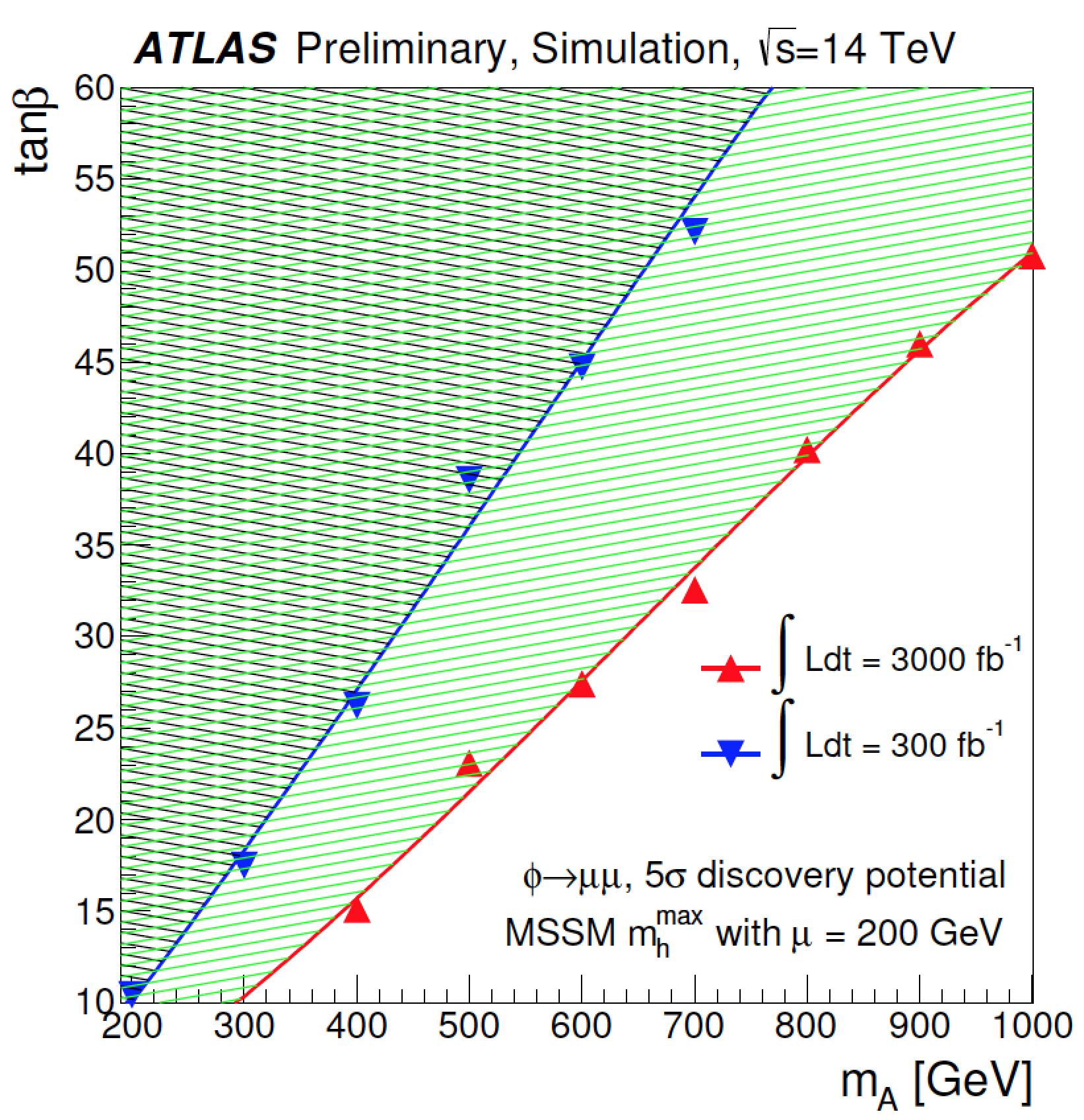}
\caption{
\label{fig:2hdm-phi-mumu}
Expected $5\sigma$ reach in the $m_A$ vs.\ $\tan\beta$ plane for the $m_h^\mathrm{max}$ MSSM benchmark scenario for the ATLAS~\cite{ATL-PHYS-PUB-2013-016}
$A/H \rightarrow \mu^+\mu^-$ analysis.
 }
\end{figure}

The $5\sigma$ reach for both categories combined is shown in Figure~\ref{fig:2hdm-phi-mumu}.

\section{Top quark decays involving Flavour Changing Neutral
  Currents}\label{sec:top-fcnc-decays}

In the standard model, where flavour changing neutral currents are not present at tree level,
top decays to a charm quark and a Higgs boson are expected to have a branching ratio of $10^{-15} - 10^{-13}$~\cite{AguilarSaavedra:2004wm,Mele:1998ag}
which is below the experimental sensitivity. 
However, several extensions of the standard model predict a branching ratio enhanced by several orders
of magnitude, including type~III two Higgs doublet models (in which $t c h$ couplings
exist at tree level) where this branching ratio can be as high as $\approx 10^{-3}$~\cite{AguilarSaavedra:2004wm}.

ATLAS~\cite{ATL-PHYS-PUB-2013-012} has estimated the sensitivity at HL-LHC to decays $t \rightarrow c h$
based on the Run I analysis~\cite{ATLAS-CONF-2013-081}. 
The selection focuses on the (very clean)
decays of Higgs bosons to a pair of photons (despite the small branching ratio).

First two photons with transverse energy greater than 40 (30) \GeV\ for the leading (subleading) photon
are required. The next steps reflect the possible decay modes of the $\mathrm{W}$ boson in the other top decay:
  
\begin{itemize}
\item {\bf hadronic $\mathrm{W}$ decays:} no charged lepton must be identified in the event and at least four jets
with $p_T > 25\ (30)\ \GeVc$ in the central (non-central)
region must be present.

\item {\bf leptonic $\mathrm{W}$ decays:} 
exactly one charged lepton is required which must have a transverse momentum above 20 \GeVc.
At least two jets above 25 (30) \GeVc\ transverse momentum in the central (non-central)
region must be present. The transverse mass, formed by the missing transverse energy
and the lepton must be larger than 30 \GeV. The longitudinal component
of the neutrino momentum is estimated from the missing transverse
energy and the charged lepton momentum imposing a $\mathrm{W}$ mass constraint.

\end{itemize}

In both channels, at least one jet must be identified as a b-jet.
To ensure consistency with the production of a pair of top quarks, 
both top decays must have reconstructed top masses close to the nominal 
top mass for one assignment of jets to the two top decays.

The expected limit on the branching ratio $t \rightarrow c h $ after $3000\ \mathrm{fb}^{-1}$, obtained
from counting the number of events for which $123 \le m_{\gamma\gamma} \le 129\ \GeVcc$,
is $1.5 \cdot 10^{-4}$ at 95\% CL. For comparison, the strictest observed limit after LHC Run I
is 0.56\%~\cite{CMS-PAS-HIG-13-034}.


\section{Conclusions and Outlook}\label{sec:conclusions-outlook}

We summarised ATLAS and CMS studies of the sensitivity to beyond the standard model Higgs physics after $3000\ \mathrm{fb}^{-1}$
in three areas:
anomalous quartic couplings in vector boson scattering, generic two Higgs doublet models
and flavour changing top decays via Higgs bosons. 
To make the estimates more 
realistic, these studies will have to be updated using a detailed simulation
of the upgraded detectors with complete simulated events and make use of reconstruction algorithms which have been
re-optimised to deal with the increased pileup at HL-LHC.

\bibliographystyle{unsrt}
\bibliography{proceedings}

\begin{thebibliography}{10}

\bibitem{Aad:2012tfa}
ATLAS Collaboration,
\newblock {Observation of a new particle in the search for the Standard Model
  Higgs boson with the ATLAS detector at the LHC}.
\newblock {\em Phys.Lett.}, B716:1--29, 2012.

\bibitem{Chatrchyan:2012ufa}
Serguei Chatrchyan et~al.
\newblock {Observation of a new boson at a mass of 125 GeV with the CMS
  experiment at the LHC}.
\newblock {\em Phys.Lett.}, B716:30--61, 2012.

\bibitem{Yang:2012vv}
Daneng Yang, Yajun Mao, Qiang Li, Shuai Liu, Zijun Xu, et~al.
\newblock {Probing W$^+$W$^-$gamma Production and Anomalous Quartic Gauge Boson
  Couplings at the CERN LHC}.
\newblock {\em JHEP}, 1304:108, 2013.

\bibitem{ATL-PHYS-PUB-2013-006}
ATLAS Collaboration,
{Studies of Vector Boson Scattering And Triboson Production with an Upgraded
  ATLAS Detector at a High-Luminosity LHC}.
\newblock Technical Report ATL-PHYS-PUB-2013-006, CERN, Geneva, Jun 2013.

\bibitem{CMS-PAS-FTR-13-006}
{Vector Boson Scattering and Quartic Gauge Coupling Studies in WZ Production at
  14 TeV}.
\newblock Technical Report CMS-PAS-FTR-13-006, CERN, Geneva, 2013.

\bibitem{Aad:2014zda}
ATLAS Collaboration,
\newblock {Evidence for Electroweak Production of $W^{\pm}W^{\pm}jj$ in $pp$
  Collisions at $\sqrt{s}=8$ TeV with the ATLAS Detector}.
\newblock 2014.

\bibitem{CMS-PAS-SMP-13-015}
{Vector boson scattering in a final state with two jets and two same-sign
  leptons}.
\newblock Technical Report CMS-PAS-SMP-13-015, CERN, Geneva, 2014.

\bibitem{Degrande:2012wf}
Celine Degrande, Nicolas Greiner, Wolfgang Kilian, Olivier Mattelaer, Harrison
  Mebane, et~al.
\newblock {Effective Field Theory: A Modern Approach to Anomalous Couplings}.
\newblock {\em Annals Phys.}, 335:21--32, 2013.

\bibitem{Eboli:2006wa}
O.J.P. Eboli, M.C. Gonzalez-Garcia, and J.K. Mizukoshi.
\newblock {$p p \rightarrow j j e^\pm \mu^\pm \nu\nu$ and $j j e^\pm \mu^\mp
  \nu \nu$ at ${\cal O}(\alpha_\mathrm{em}^6)$ and ${\cal
  O}(\alpha_\mathrm{em}^4 \alpha_s^2)$ for the study of the quartic electroweak
  gauge boson vertex at CERN LHC}.
\newblock {\em Phys.Rev.}, D74:073005, 2006.

\bibitem{Eriksson:2009ws}
David Eriksson, Johan Rathsman, and Oscar Stal.
\newblock {2HDMC: Two-Higgs-Doublet Model Calculator Physics and Manual}.
\newblock {\em Comput.Phys.Commun.}, 181:189--205, 2010.

\bibitem{ATL-PHYS-PUB-2013-016}
ATLAS Collaboration,
{Beyond Standard Model Higgs boson searches at a High-Luminosity LHC with
  ATLAS}.
\newblock Technical Report ATL-PHYS-PUB-2013-016, CERN, Geneva, Oct 2013.

\bibitem{CMS-PAS-FTR-13-024}
{2HDM Neutral Higgs Future Analysis Studies}.
\newblock Technical Report CMS-PAS-FTR-13-024, CERN, Geneva, 2013.

\bibitem{Aad:2013wqa}
ATLAS Collaboration,
\newblock {Measurements of Higgs boson production and couplings in diboson
  final states with the ATLAS detector at the LHC}.
\newblock {\em Phys.Lett.}, B726:88--119, 2013.

\bibitem{deFavereau:2013fsa}
J.~de~Favereau et~al.
\newblock {DELPHES 3, A modular framework for fast simulation of a generic
  collider experiment}.
\newblock {\em JHEP}, 1402:057, 2014.

\bibitem{AguilarSaavedra:2004wm}
J.A. Aguilar-Saavedra.
\newblock {Top flavor-changing neutral interactions: Theoretical expectations
  and experimental detection}.
\newblock {\em Acta Phys.Polon.}, B35:2695--2710, 2004.

\bibitem{Mele:1998ag}
B.~Mele, S.~Petrarca, and A.~Soddu.
\newblock {A New evaluation of the t $\rightarrow$ cH decay width in the
  standard model}.
\newblock {\em Phys.Lett.}, B435:401--406, 1998.

\bibitem{ATL-PHYS-PUB-2013-012}
ATLAS Collaboration,
{Sensitivity of ATLAS at HL-LHC to flavour changing neutral currents in top
  quark decays t $\rightarrow$ cH, with H $\rightarrow \gamma \gamma$}.
\newblock Technical Report ATL-PHYS-PUB-2013-012, CERN, Geneva, Sep 2013.

\bibitem{ATLAS-CONF-2013-081}
ATLAS Collaboration,
{Search for flavor-changing neutral currents in $t\rightarrow cH$, with
  $H\to\gamma\gamma$, and limit on the tcH coupling}.
\newblock Technical Report ATLAS-CONF-2013-081, CERN, Geneva, Jul 2013.

\bibitem{CMS-PAS-HIG-13-034}
{Combined multilepton and diphoton limit on t to cH}.
\newblock Technical Report CMS-PAS-HIG-13-034, CERN, Geneva, 2014.

\end{thebibliography}

\end{document}